\documentclass[aps,prl,twocolumn,superscriptaddress]{revtex4}
\usepackage{graphicx,psfrag}
\usepackage[latin1]{inputenc}
\usepackage{natbib}
\usepackage{amsmath, amsthm, amssymb}

\newcommand{\notes}[1]{}

\newcommand{\beq}{\begin{equation}}
\newcommand{\eeq}{\end{equation}}
\newcommand{\beqnn}{\begin{equation*}}
\newcommand{\eeqnn}{\end{equation*}}
\newcommand{\beqas}{\begin{eqnarray*}}
\newcommand{\eeqas}{\end{eqnarray*}}
\newcommand{\beqa}{\begin{eqnarray}}
\newcommand{\eeqa}{\end{eqnarray}}

\begin{document}

%\title{Spin wave oscillators: Towards a new computing paradigm}
%\title{Frequency Analysis of Spin Wave Generators}% this is only a suggestion....
\title{Spin reversal in Fe$_8$ under fast pulsed magnetic fields}

\author{S. N. Jammalamadaka}
\affiliation{Pulsed Field Group, Institute for Nanoscale Physics and Chemistry (INPAC), KU Leuven, Celestijnenlaan 200D, B-3001, Leuven, Belgium}
\affiliation{Department of Physics, Indian Institute of Technology Hyderabad, Ordnance Factory Estate, Yeddumailaram, Andhra Pradesh, India, 502205}
\author{J. Vanacken}
\affiliation{Pulsed Field Group, Institute for Nanoscale Physics and Chemistry (INPAC), KU Leuven, Celestijnenlaan 200D, B-3001, Leuven, Belgium}
\author{V. V. Moshchalkov}
\affiliation{Pulsed Field Group, Institute for Nanoscale Physics and Chemistry (INPAC), KU Leuven, Celestijnenlaan 200D, B-3001, Leuven, Belgium}
\author{\\S. V\'elez}
\affiliation{Grup de Magnetisme, Departament de F\'isica Fonamental, Universitat de Barcelona, Barcelona 08028, Spain}
\affiliation{CIC nanoGUNE, 20018 Donostia-San Sebastian, Basque Country, Spain}
\author{J. Tejada}
\affiliation{Grup de Magnetisme, Departament de F\'isica Fonamental, Universitat de Barcelona, Barcelona 08028, Spain}
\author{F. Maci\`a}
\email{ferran.macia@ub.edu}
\affiliation{Grup de Magnetisme, Departament de F\'isica Fonamental, Universitat de Barcelona, Barcelona 08028, Spain}

\date{\today}

\begin{abstract}
We report measurements on magnetization reversal in the Fe$_8$ molecular magnet using fast pulsed magnetic fields of 1.5 kT/s and in the temperature range of 0.6-4.1 K. We observe and analyze the temperature dependence of the reversal process, which involves in some cases several resonances. Our experiments allow observation of resonant quantum tunneling of magnetization up to a temperature of $\sim$ 4 K. We also observe shifts of the resonance fields in temperature that suggest the emergence of a thermal instability---a combination of spin reversal and self-heating that may result in a magnetic deflagration process. The results are mainly understood in the framework of thermally-activated quantum tunneling transitions in combination with emergence of a thermal instability.
\end{abstract}

\maketitle

\section{Introduction}

Single molecule nanomagnets have been of great interest because of their quantum effects \cite{Tejada1998,JonathanPRL1996,Hernandez1996,Thomas1996,Sessoli1999,chudnovsky1997,Werns2000} and possible applications in quantum computers \cite{Loss2001,tejada_Nanotech} or magnetic refrigerants \cite{torres2000}. Spin-level populations in nanomagnets are easily to manipulate by modifying their energy with external magnetic fields. Among the large set of synthesized molecule magnets, Mn$_{12}$ and Fe$_8$ have been the most studied because of their relatively easy preparation, large molecular spin, and large magnetic anisotropy. Dynamics of spin at low temperatures had been studied---including quantum tunneling magnetization (QTM)\cite{JonathanPRL1996} and electron paramagnetic resonance (EPR)\cite{HillEPR,SessoliEPR}---and has been described through the so-called Giant Spin Approximation (GSA) \cite{Tejada1998} that assigns a single spin quantum number, $S$, to the ground-state spin levels. Effectively, the spins relax toward equilibrium through a combination of thermal activation and quantum tunneling \cite{JonathanPRL1996,chudnovsky1997}.
\\
\\
Fe$_8$ was initially prepared by Wieghardt \emph{et al.} \cite{Weighardt} and has shown clear evidence of QTM \cite{Sangregorio1997,Sessoli1999,wernsJAP2000,Werns2000,SessoliAngew2003}. At low temperatures, the eight iron cations assemble couple in such a manner that give rise to a high spin $S=10$ molecule with an anisotropy barrier height of about 29 K. The spin Hamiltonian for the Fe$_8$ is given by \cite{barra1996,Sangregorio1997}
\beq
\label{ham}
\mathcal{H} = - DS_z^2 + E(S_x^2-S_y^2) + \mathcal{H}_{\text{ho}} - g\mu_B\ H_zS_z\ ,
\eeq
where anisotropy constants $D$ and $E$ are 0.292 and 0.046 K respectively and $g \approx 2$. Both anisotropies have been extensively measured through high frequency-EPR \cite{barra1996} and neutron spectroscopy \cite{Caciuffo_INS}. The first term in the Hamiltonian defines the anisotropy barrier and creates an easy axis ($z$-axis) for the magnetization. The second and third terms break the rotational symmetry of the Hamiltonian and are responsible for the tunneling of the magnetization. The term $\mathcal{H}_{\text{ho}}$ stands for high order terms. The last term of the Hamiltonian describes the Zeeman energy associated with an applied field $H_z$ in the direction of the easy axis.
\\
\\
In this Letter we investigate the temperature dependence of the magnetization reversal in the Fe$_8$ molecular magnet under fast pulsed magnetic fields of 1.5 kT/s. This fast sweep rate allows observation of quantum tunneling of the magnetic moment at high temperatures up to $\sim$ 4 K. The measurements of spin dynamics are modeled with thermally-activated quantum tunneling transitions accounting for the self-heating that might result in a magnetic deflagration \cite{subedi2013}

\section*{Experimental set-up}
\begin{figure}[htb]
\includegraphics[width=\columnwidth]{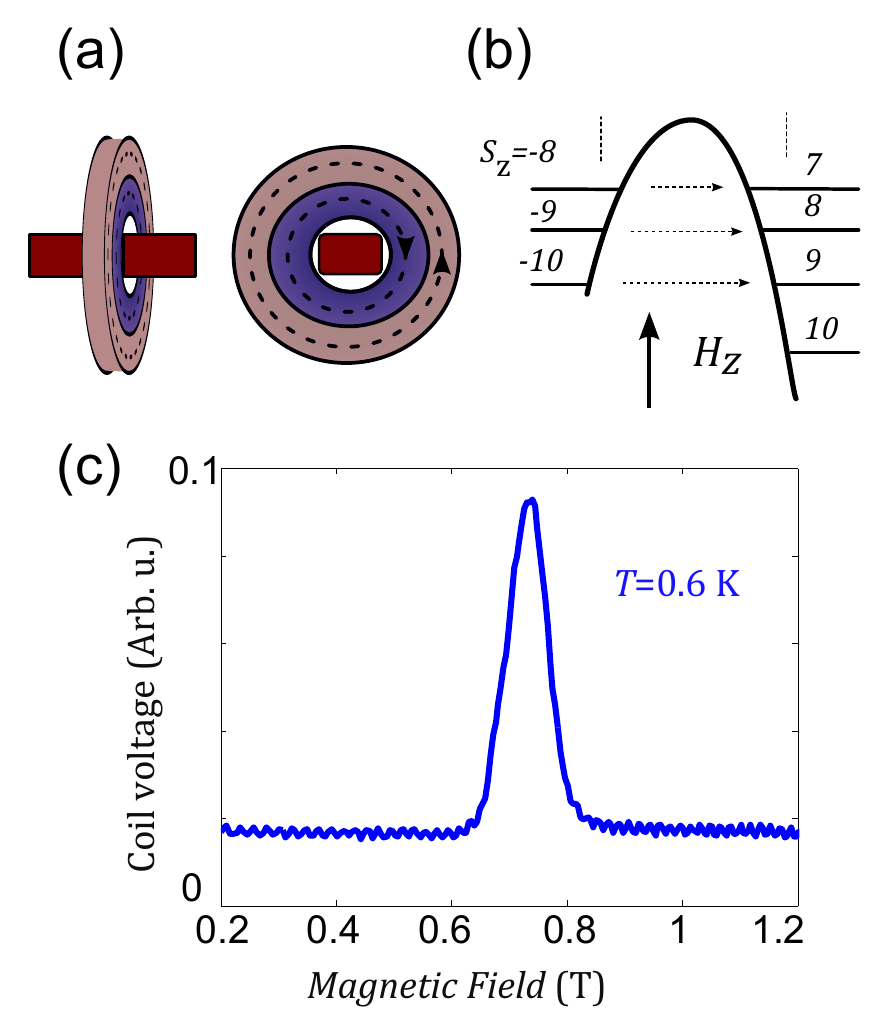}
\caption{\small{(a) Schematic of the compensated coil susceptometer. (b) Energy level diagram (two wells) described by the Hamiltonian (Eq. 1) (c) Voltage trace recorded during magnetization reversal while applying a magnetic pulse that produce a linear sweep rate of 1.5 kT/s. Once the magnetization of sample has reversed, new pulses in the same field polarity do not show any variation in the coil voltage signal.}}
\label{fig1}
\end{figure}

A single crystal of Fe$_8$ molecular magnet with dimensions aproximately 1$\times$0.15$\times$0.27 mm  was used for the present investigation. The sample was embedded in a cylinder form in stycast (but it did not change its plate-like shape). Pulsed magnetic field measurements of 1.5 kT/s were performed at the Pulsed Fields facility of the Katholic University of Leuven, Belgium. A coil with an inductance of 650 $\mu$H was used to generate magnetic field pulse. This facility allowed us to go up to 70 T with pulse duration of 20 ms by discharging the capacitor bank through a specially designed magnet coil \cite{Vanacken}. %Different field sweep rates were obtained by tuning the capacitances of capacitor bank from $C = 2$ mF to $C = 38$ mF while the voltage was altered from $V = 5000$ V to $V = 600$ V in such a way that the capacitor energy, $1/2CV^2$, remained constant.
Magnetization reversal was detected using a compensated coil (See Fig.~\ref{fig1}a), which was done by tuning the number of inner and outer windings.  The compensated coil was sensitive to the signal coming from the sample and insensitive to the signal coming from the applied pulses. Low temperatures down to 0.6 K were achieved using a He$^3$  cryostat. The sample holder was made of non-metallic materials and was submerged in liquid He$^3$ during the measurements. The temperature was varied in the range between 0.6 - 4.1 K by pumping the helium with control through a needle valve.

The crystal was initially magnetized to negative saturation at a fix bath temperature. Then we swept the magnetic field at an ultra fast rate of 1.5 kT/s, so the Fe$_8$ crystal is in a non-equilibrium magnetization state, as shown in Fig.\ \ref{fig1}(b). We recorded the signals of the compensated coil during the complete field sweep. The magnitude of the peak voltage captured with the coils, $V$, is proportional to the magnetic flux variation, $dB/dt$, being therefore $V\propto dM/dt$ of the sample and indicating the spins reversal. Figure 1(c) shows an example of the signal recorded as a function of the applied magnetic field during a magnetization reversal.

\section*{Results}

\begin{figure*}[htb]
\includegraphics[width=160mm]{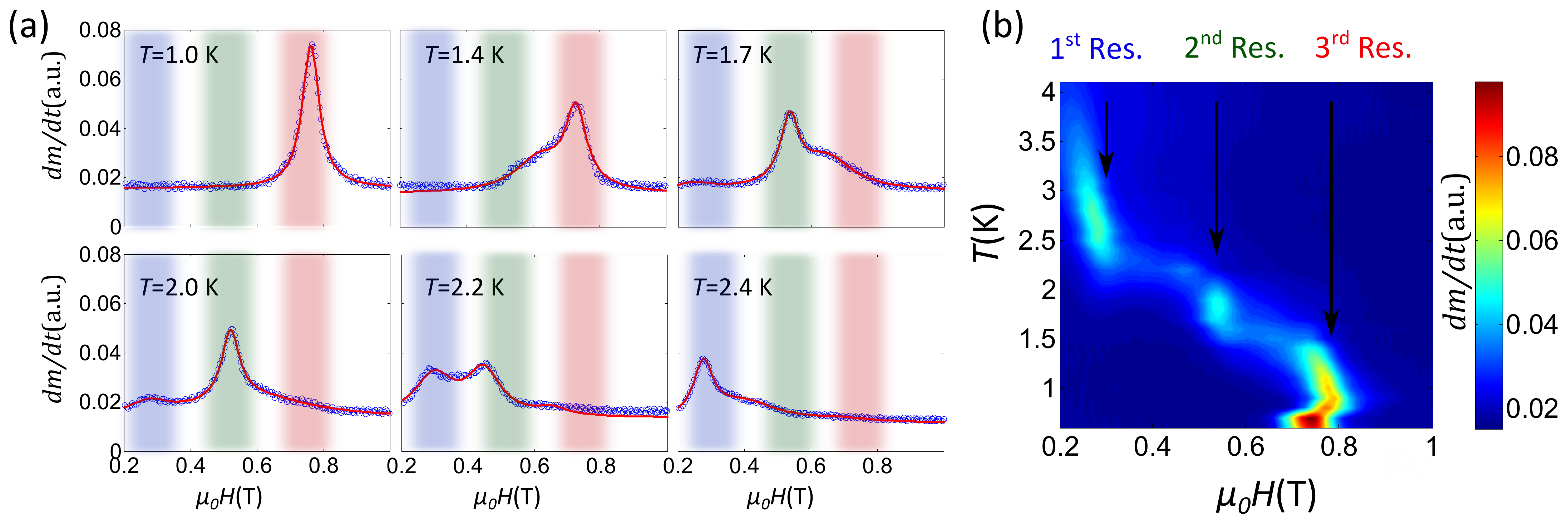}
\caption{\small{Magnetization reversal of the Fe$_8$ crystal taken at different temperatures for a fixed sweeping magnetic field rate of 1.5 kT/s. (a) shows 6 curves for the coil voltage at different temperatures. All panels have the same scale and include a fitting in red that correspond to three Lorentzian peaks. The color bands represent the field range for first (blue), second (green), and third (red) resonances. (b) corresponds to a color-scale 2d plot of all measured curves.}}
\label{fig2}
\end{figure*}

Figure\ \ref{fig2} shows the temperature dependence of the magnetization reversal for a sweeping magnetic field of 1.5 kT/s. We first plotted six representative curves at different temperatures in Fig.\ \ref{fig2}a and we fitted the curves with three Lorentzian peaks corresponding to the spin reversal at each resonance ($H_z = nH_R$ with $\mu_0H_R\sim0.26$ T and $n=1$, 2, and 3 being the order of the resonance). Each curve corresponds to the reversal of the magnetization at a fixed bath temperature. A color-scale 2d plot shows in Fig.\ \ref{fig2}b all the measured curves for temperatures between 0.6 and 4 K. We can see both the variation in amplitude and position of the peaks. Depending on temperature, the resonant field $H_z = nH_R$ changes, from the third resonance at low temperatures $T<1.2$ K to the first resonance at high temperatures $T>2.4$ K. Further, we observe that at certain temperatures, relaxation occurs throughout more than one resonance field. We notice here that the magnetization reversal peaks at 0.6 K and 0.7 K show a different trend compared to the rest of the temperatures. The resonances occur at fields larger than expected ($\mu_0H_R\sim0.26$ instead of $0.22$ T). Previous experiments at high sweep rates showed similar field shifting \cite{marta}. Although we cannot avoid certain misalignment of a few degrees we cannot attribute this large variation to that. We will discuss this in more detail later on.

\begin{figure}[htb]
\includegraphics[width=60mm]{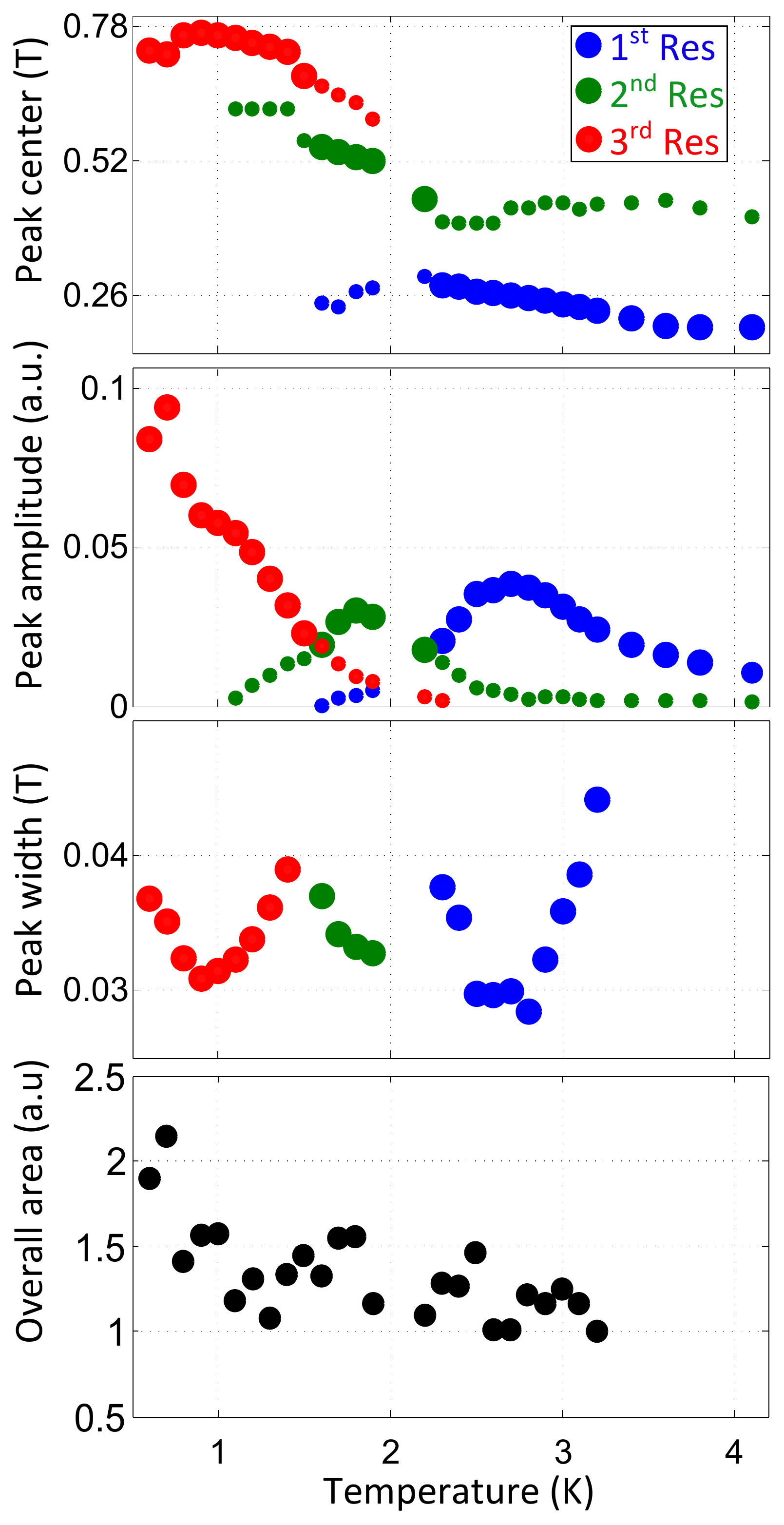}
\caption{\small{Temperature dependence of the reversing magnetization peaks shown in Fig.\ \ref{fig2}: center of the resonance peak (first panel), amplitude of the resonance peak (second panel), width of the resonance peak (third panel), and overall area under the coil's voltage (fourth panel). In the three upper panels, blue corresponds to first resonance field, green, to second, and red, to third. We have enlarged the circles  representing the peak that dominated the spin reversal process at each given temperature.}}
\label{fig3}
\end{figure}

As we increase the temperature we see how the reversal of the magnetization shifts towards lower magnetic fields while staying at the same resonant field. Peak position moves first within one resonance field and then jumps to the next smaller resonance field where it continues shifting with the same trend. The jump from the third dominant resonance field to the second one takes place around $T\sim1.6$ K whereas the drop from the second to the first occurs around $T\sim 2.2$ K. Detailed plots with peak analysis of the Lorentzian fits are displayed in Fig.\ \ref{fig3}. The top panel of Fig.\ \ref{fig3} shows the resonance field values of the reversing magnetization peaks as a function of the temperature, the second panel shows peak amplitude, and the third panel, peak width. Peak amplitudes and peak widths have a negative correlation; when one increases the other one decreases and vice versa. This observation agrees well with having a total magnetization variation being roughly constant---especially when the reversal is dominated by a single peak. However, when we plotted the overall area under the coil's voltage (see, lower panel of Fig.\ \ref{fig3}) we see that it slowly decreases with temperature suggesting the overall magnetization variation lowers with temperature. The initial magnetization state of our system once we begin measuring magnetization reversal (our measurements began at $\mu_0H=0.1$ T) might be different because some spins might have already reversed at the zero field resonance \cite{wernsJAP2000}.

%The low ($T<1.2$ K) and high ($T>2.5$ K), temperature regimes allow observation of the spin reversal while having the peaks in a same resonance, being it the third ($H\sim0.72$ T) and the first ($H\sim0.24$ T), respectively; In between the two regimes the reversal of magnetization occurs at several resonant fields and the amplitude and hight of peaks becomes more difficult to analyze.

%However, looking at the low and high temperature regimes we see how the resonant frequency slightly decreases with temperature, the peak amplitude also decreases with temperature and the peak width increases with temperature.

\section*{Discussion}

\begin{figure}[h]
\includegraphics[width=\columnwidth]{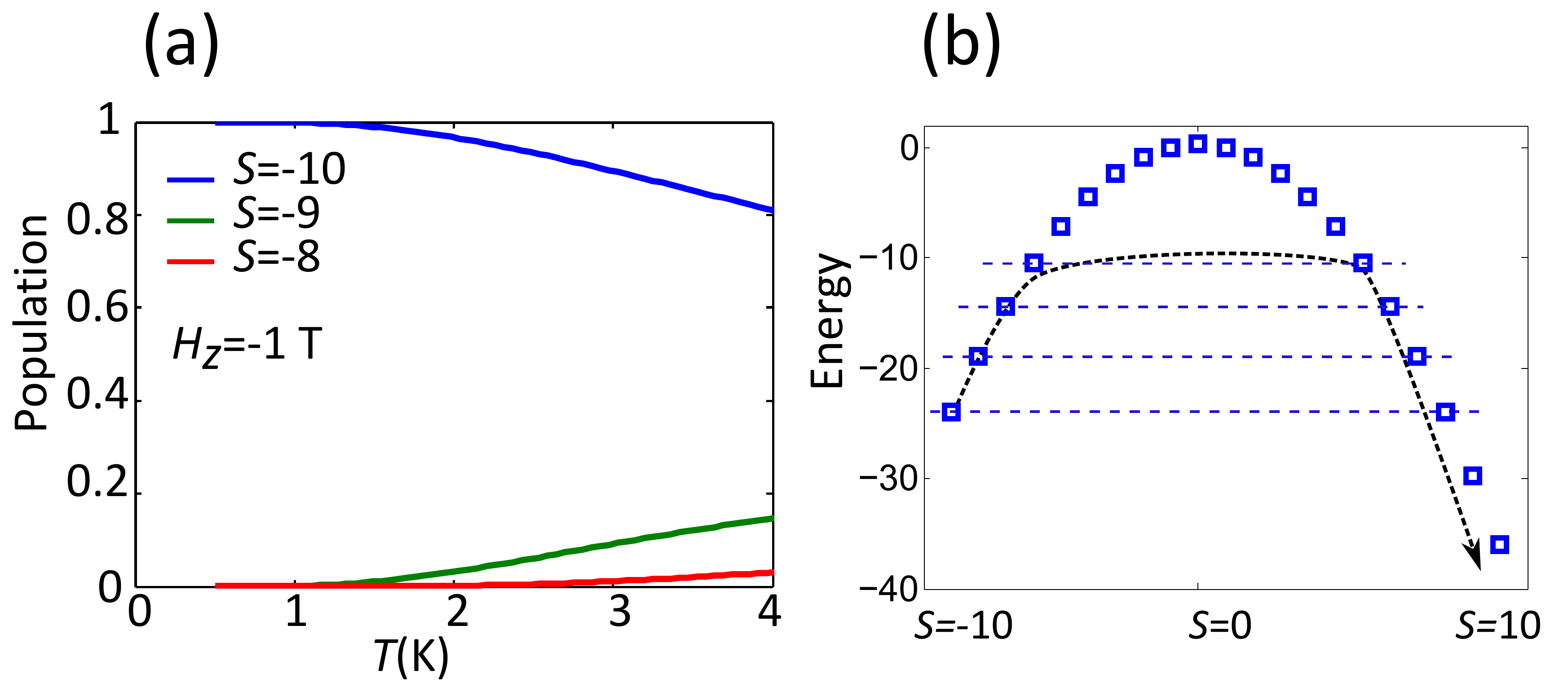}
\caption{\small{ (a) Population of the lower spin levels in the negative well at a negative field, $H=-1$ T. This configuration should correspond to the magnetic state of the system before spins begin to relax towards equilibrium during the fast magnetic field sweep. (b) Energies of the spin levels in the Fe$_8$ when the applied field corresponds to the second resonance. The dashed arrow plots the hypothetical trajectory of the spins.% (c) Temperature  dependence of the relaxation rate, $\Gamma$. Three different applied fields are plotted, $\mu_0 H=0.39, 0.44$ and 0.49 T.}}
}}
\label{Energies}
\end{figure}

To understand the temperature dependence of the fast spin reversal we need to account for all transition probabilities among levels---thermal activation, decay, and tunneling---and level populations. The population variation of levels as a function of applied field, $H_z$, and temperature, $T$, are well known. Figure\ \ref{Energies}(a) shows the population of the lower levels for a large negative field (-1 T). This population configuration is similar to the population configuration right before the spins begin to relax towards equilibrium when $H$ is swept from negative to positive values in our experiments.
A longitudinal field $H_z$ can drive the crystal in and out of tunneling resonances, reducing effectively the barrier separating spin-up and spin-down states (see, Fig.\ \ref{Energies}(b)).
%In practice, the splitting, $\Delta_{m,m'}$, of the $(m,m')$-resonance is much smaller than the widths, $\gamma_{m,m'}$, of the tunneling levels due to spin-phonon and other decay processes. Thus spins do not coherently oscillate between $m$ and $m'$ but decay to the ground state $m = S$ right after tunneling from $m$ to $m'$. The rate of the quantum transition to the stable well from level $m$ is given by \cite{garanin1997}
%
%\begin{equation}
%\Gamma_{m,m'} =
%\frac{\Delta_{m,m'}^2}{2\hbar^2}\frac{\gamma_{m'}/2}{\omega_{m,m'}^2 +
%({\gamma_{m'}/2})^2}\,,
%\end{equation}
%
%where $\hbar\omega_{mm'} = \epsilon_m - \epsilon_{m'}$ is the detuning of the $m$-th and $m'$-th levels.

The computation of the splitting $\Delta_{m,m'}$ of the $(m,m')$-resonance, in a given sample is not easy because the crystal is a macroscopic object that has demagnetizing fields both $h_x$ and $h_z$ that depend on the crystal shape and the overall tunneling is very sensitive to the transverse applied fields and it has a parity effect \cite{Sessoli1999}. For instance, in absence of applied $h_x$, a single molecule will only have tunneling at the even resonances ($n$ = 0, 2, 4 \dots). However, experimentally it is observed that both even and odd resonances show quantum tunneling. The presence of transverse fields varies critically, mixing among spin states, and allows tunneling between odd resonance levels \cite{Sessoli1999}. In our experiments we have mentioned that there might be a misalignment, mainly because the crystallographic structure of Fe$_8$. For instance, a small misalignment of 7 degrees would produce a transverse magnetic field of 0.1 T at the third resonance.

In order to estimate values for the level splittings we calculate a simple case where there are no transverse fields (i.e., we have to compute even resonances). Using perturbation theory \cite{Tejada1998} the values for the rate of quantum transition, $\Gamma_{m,m'}=\frac{\Delta_{m,m'}^2}{2\Gamma_0\hbar^2}$ with $\Gamma_0=\tau^{-1}\sim10^{7}$ being the characteristic time of the system, in Hz for the second resonance ($H_z=2H_R$) %at zero detuning, and considering the level's width be $\gamma_{m,m'}=\Gamma_0=10^7$ Hz, we obtain the following values for the second resonance:
\beqnn
\begin{array}{ll}
\Gamma_{-10,8}=3.6\cdot 10^{-10},&\Gamma_{-9,7}= 3.5\cdot 10^{-5},\\
 \Gamma_{-8,6}=8.3\cdot 10^{-1}, & \Gamma_{-7,5}=6.4\cdot10^{3},\\
 \Gamma_{-6,4}=1.7\cdot 10^{7}.&
\end{array}
\eeqnn

Once $\Gamma_{mm'}$ exceeds the thermally activated relaxation rate from the $m$-th level, the barrier is effectively reduced due to underbarrier tunneling from the $m$-th level (see, Fig.\ \ref{Energies}(b)). Spin reversal in our experiments occurs in about 100 $\mu$s (the sweeping field varies 0.15 T in 100 $\mu$s) that corresponds to a relaxation rate, $\Gamma$, of the order of 10 kHz. We notice that the first quantum transition that would allow some variation of the magnetization is $\Delta_{-7,5}$ (and  $\Delta_{-6,4}$ for a complete), indicating that the effective barrier would be reduced to about 15 K.

%We observed spin reversal occurring in about 100 $\mu$s (the sweeping field varies 0.15 T in 100 $\mu$s) and the first rate of quantum transition that would allow spin reversal would be $\Delta_{-7,5}$.

%We have observed spin reversal occurring in about 100 $\mu$s (the sweeping field varies 0.15 T in 100 $\mu$s) and the first rate of quantum transition that would allow the reversal of spins would be $\Delta_{-7,5}$. Additionally, spins need to be activated thermally to reach, at least, level $m=-7$ in the metastable well, which corresponds to about 13 K for the second resonance, (see Fig.\ \ref{Energies}b). In this case, the equivalent problem is to consider a barrier height of 13 K and therefore the minimum temperature needed to allow reversal of the magnetization in 100 $\mu$s would be about 2 K. %This calculations indicate that the spin reversal must occur at slightly larger temperatures than the measured bath temperatures suggesting there might be self heating during the spin reversal.

In addition, our experiments show a variation of the resonance field within the same-resonance level that cannot be explained by the high-order terms in the Hamiltonian (Eq.\ \ref{ham}). Spin levels in Fe$_8$ molecule resonate all at the same applied field and variations in temperature (i.e., variations in the spin-level population) should not vary the field at which spin tunneling occurs. We believe that the observation of the shifting corresponds to the emergence of a thermal instability---that may result in a magnetic deflagration---related with the self-heating process due to fast reversal of magnetization. In the following paragraphs we compare slow relaxation against ignition of a deflagration due to a thermal instability.

% Higher-order terms ($\mathcal{H}_{\text{ho}}$) in the spin Hamiltonian, Eq.\ \ref{ham}, such as $S_z^4$ would cause that resonance levels have a non-even spacing and therefore tunneling at different levels would occur at slightly different fields (this is the case of the Mn$_{12}$ molecule magnet \cite{mertes2002}). Higher-order terms were actually measured \cite{Caciuffo_INS} in Fe$_8$ crystals and values that were about 4 orders of magnitude smaller than $D$ (from Eq.\ \ref{ham}) were reported for the $S_z^4$. However, a fourth order term of at least one order of magnitude higher than the measured one \cite{Caciuffo_INS} would be necessary to explain the observed shifting of the resonance fields.
%%% WE HAD THIS HERE IN THE FIRST RESUB VERSION AND NOW WE MOVED IT TO THE END

%%%%%here we go with deflagration
Magnetic deflagration has been extensively studied in Mn$_{12}$-ac \cite{suzuki,quantumdeflagration,Garanin2007,subedi2013}, including experiments at high-field sweep rates \cite{Decelle}. In a magnetic relaxation process there is a competition between heat produced by the reversing spins and heat diffused throughout the sample and the bath; if the diffusion cannot compensate the reaction term, the process becomes unstable and some magnetic materials \cite{suzuki,macia2007,velez2010} experience a transition between thermal relaxation and a fast-propagating spin-reversal process named magnetic deflagration \cite{subedi2013}.

The dynamics of the magnetization system when relaxing could be described by the following equations
\beq
\dot{m}=-\Gamma(m-m_{\text{eq}}), \qquad \dot{T}=\dot{m}\Delta E/C+\nabla \cdot  \kappa \nabla T,
\label{inst1}
\eeq
where $\Gamma=\tau^{-1}$ is the relaxation rate, $m_{\text{eq}}$ is the equilibrium magnetization, $\Delta E$ the energy released, $C$ the heat capacity and $\kappa$ the thermal diffusivity. The condition for the system to lose stability is given by
\beq
\dot{T}=0, \qquad \partial{\dot{T}}/\partial{T} = 0.
\label{inst2}
\eeq

According to the theory of magnetic deflagration \cite{Garanin2007,MaciaHill}, the threshold for ignition of the deflagration process is achieved when the rate, $\Gamma$, of the transition out of the metastable well exceeds a critical value (this follows from Eqs.\ \ref{inst1} and \ref{inst2})
\begin{equation}
\Gamma _{c}=\frac{8\kappa k_{B}T_{0}^{2}}{U_{\text{eff}}\Delta E\,n_{i}l^{2}} ,
\label{threshold}
\end{equation}
where $U_{\text{eff}}(H)$ and $\Delta E(H)$ are the field-dependent effective energy barrier and the energy difference between spin-up and spin-down ground states, respectively; $l$ is a characteristic length of the order the smallest dimension of the sample; and $n_{i}$ is the initial population in the metastable well. Once the relaxation rate is larger than the critical value given in Eq.\ \ref{threshold} it means that the heat produced by the reversing spins cannot be compensated by the diffusion and therefore the sample's temperature will increase resulting in a faster spin relaxation and a faster heat production. Eventually, the magnetization reverses completely and the sample's temperature cools down to the bath temperature.

In our experiment, the magnetization reverses in less than 100 $\mu$s---that corresponds to a spin relaxation rate of $\Gamma=$10 kHz. However, if we calculate the corresponding threshold for the relaxation rate that would induce a magnetic deflagration (see Eq.\ \ref{threshold}) at the second resonance for a temperature of $T=2$ K, we obtain $\Gamma_c=20$ Hz$-$2 kHz depending the value for the thermal diffusivity, ($\kappa=10^{-6}-10^{-4}$). This means that the thermal instability that could lead to a magnetic deflagration would originate before the relaxation rate could increase to the value observed in our experiments ($\Gamma=10$ KHz). Reported experiments in Fe$_8$ at much slower field sweep rates \cite{wernsJAP2000} showed lower relaxation rates, on the order of the second, and thus the deflagration condition was not fulfilled.

%The first rates of quantum transition that would allow spin reversal would be $\Delta_{-7,5}$ and $\Delta_{-6,4}$.

%%%%%%%%%%%%%%%%%%%%%

%We now compare the stability against slow relaxation (meaning a relaxation process that do not produce a considerable self-heating) given by the condition $\Gamma(H,T_0) t > 1$, with $t$ being the characteristic time of the experiment, with the deflagration condition given by Eq.\ \ref{threshold}. We consider the following case: \emph{i)} applied fields around the second resonance ($\mu_0H_R\approx 0.22$ T) \emph{ii)} an effective barrier, $U_{\text{eff}}$ as the energy difference between the ground state, $m=-10$, and the first level where the tunnel rate is faster than 1 ms (level $m=-7$) \emph{iii)} a detuning of about 0.05T (i.e., there is tunneling in the interval $\mu_0H_R\pm0.05$ T).

%We mentioned we do not expect a field shift within the resonance field as a function of the temperature if the reversing process occurs with no self-heating.
The ignition of a thermal instability is sensitive to the bath temperature (see, Eq.\ \ref{threshold}). Once the system reaches the instability---and spins begin to reverse and temperature, to increase---there is still a waiting time until the deflagration front forms or, in other words, until the temperature rises enough to allow a spin-reversal at the timescale of our sweeping field. This waiting period is sometimes called \emph{ignition time} \cite{quantumdeflagration}.

\begin{figure}[h]
\includegraphics[width=\columnwidth]{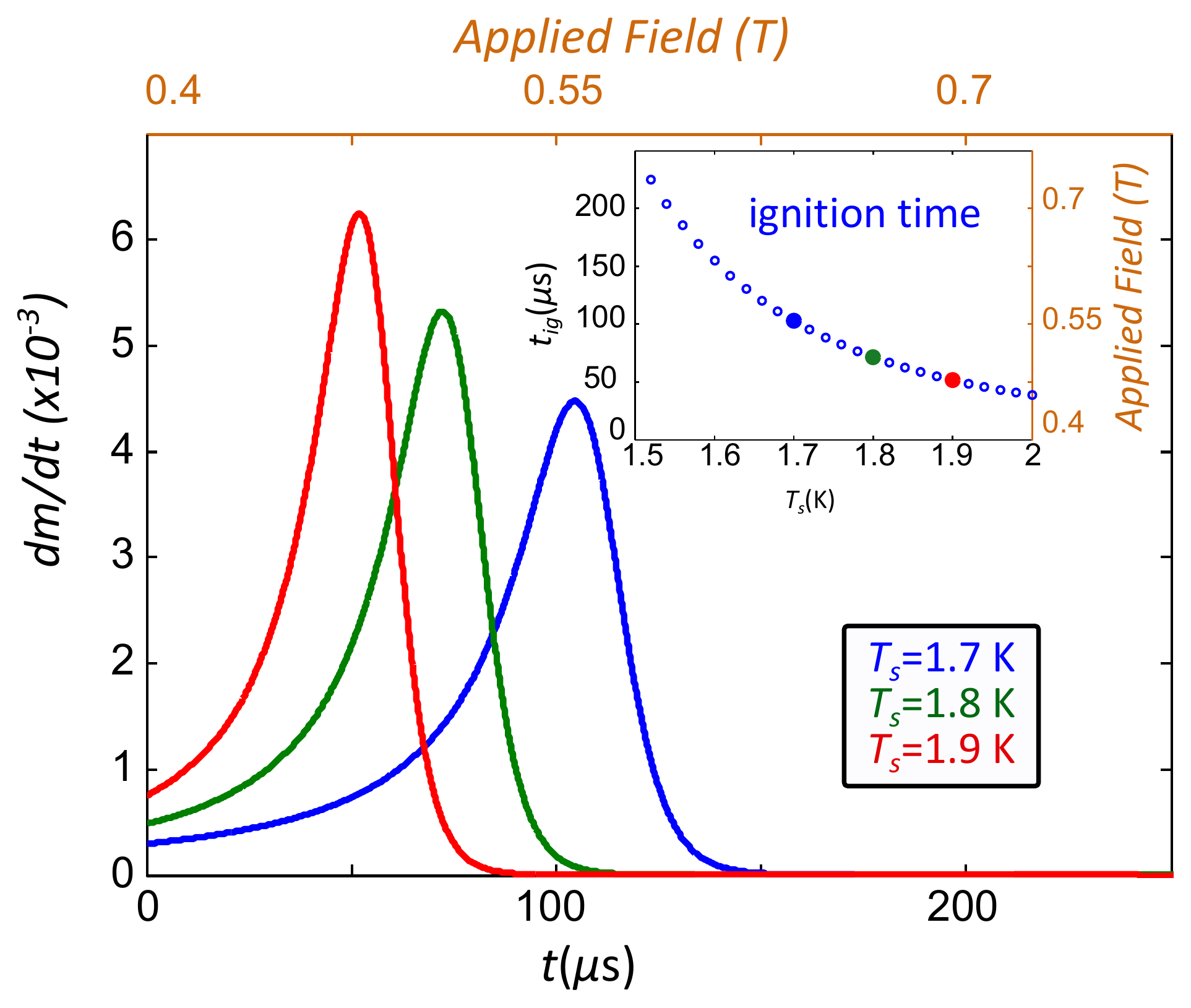}
\caption{\small{ Time derivative of the magnetization evolution of the spin reversal following Eq.\ \ref{1d} for different bath temperatures, $T_s$=1.7, 1.8, and 1.9 K. Top $y$-axis is labeled with with applied magnetic field assuming a sweeping rate as in the experiment of 1.5 kT/s ($t=0$ corresponds arbitrarily to $\mu_0H=0.4$ T). The inset shows the bath temperature dependence of the ignition time defined as the peak of main panel curves at each temperature. The right-hand-side axis is labeled as well with the applied field.}}
\label{fig5}
\end{figure}

Next we compute the ignition times as a function of the bath temperature for the case described in Fig.\ \ref{Energies}b (second resonance) where we consider an effective barrier of 15 K. Let us take here the magnetization and temperature evolution in a nucleation volume, independent of coordinates, as described in Eq.\ \ref{inst1}
\beq
\dot{m}=-\Gamma(m-m_{\text{eq}}), \qquad \dot{T}=\dot{m}\Delta E/C+ \frac{\kappa}{2l^2}(T-T_b),
\label{1d}
\eeq
where $T_b$ is the bath temperature and $2l$ is the characteristic size of the nucleation volume and is bounded by the smallest sample dimension (0.15 mm). The diffusive term in Eq.\ \ref{1d} is linear with temperature while the reaction term increases exponentially with temperature ($m\propto \exp[-U/(k_BT)]$). The competition between this two terms sets the critical value, $\Gamma_c$, for the thermal instability. We have computed magnetization and temperature evolutions from Eq.\ \ref{1d} at different bath temperatures. Figure \ref{fig5} shows representative time derivatives of the magnetization curves at different bath temperatures, $T_b$, using a heat capacity close to the measured value $C/(k_BN_A) \sim 1$ [i.e., $C=$ 8.3 J/(mol K)] \cite{PhysRevLett.93.117202} and a thermal diffusivity of $\kappa=10^{-5}$. The curves shown correspond to cases where the instability condition, $\Gamma>\Gamma_c$, is reached and thus the temperature increases and the magnetization reversal accelerates. We see that as the temperature increases the reversal of magnetization occurs earlier---because the thermal instability develops faster.

%We see that at larger bath temperatures the magnetization reversal develops faster---the represented curves in Fig.\ \ref{fig5} show a larger amplitude and a smaller width at lower temperatures. We defined the ignition time as the maximum of the magnetization variation for each computed curve and we plotted it in the inset of Fig.\ \ref{fig5}. %The dependence of the ignition time as a function of the initial temperature decays almost exponentially.

The ignition time depends strongly on the bath temperature and in our experiment, where field varies at 1.5 kT/s, this results in a shift of the field where we detect the spin reversal. We labeled the top $y$-axis of the Fig.\ \ref{fig5} main's panel and the right-hand-side axis of its inset with the applied field that will correspond to a given delay accounting for the used sweeping rate of 1.5 kT/s (we took $t=0$ for the field $\mu_0H=0.4$ T). The magnetization reversal processes may start as soon as the field reaches the resonance, and depending on the bath temperature, the deflagration takes more, or less, time to develop and consequently the spin reversal occurs at a different field. We notice that the computed curves have a width of 50-100 $\mu$s that corresponds to 75-150 mT, which agrees well with experimental curves. Additionally, the ignition time is of the order of tens of microseconds indicating that all the measured curves might be shifted towards larger fields (10 $\mu$s corresponds to 15 mT), giving also an explanation for the unexpected shift observed at the resonance-field values, rather than a large misalignment.

Finally, we notice that the lowest measured temperatures (0.6-0.8 K) show a different trend in the magnetization reversal when comparing it with all other higher temperatures. We were not able to measured other curves in this regime and with the present data we are unable to describe the origin of this anomaly. However, we are aware that this feature may be of enormous interest because one might enter into rather unknown regimes for the Fe$-8$ molecular magnet such as collective electromagnetic emission \cite{Eugene2002_superR,marta,keren} or magnetic detonation \cite{Decelle}.

%but we would like to suggest the possibility that the spin reversal process enters into a different regime. Previous experiments have shown that magnetic relaxation could be mediated by collective electromagnetic emission in Fe${_8}$ at $T = 0.6$ K when a magnetic field is swept at high field rates \cite{Eugene2002_superR,marta,keren}, and this would cause entering into a different regime for the magnetization dynamics.

In conclusion we have measured spin reversal through quantum tunneling at temperatures up to $\sim 4$ K in the Fe$_8$ molecular magnet using ultra fast pulsed magnetic fields of 1.5 kT/s. Measuring with high-fields sweep rates allowed us the observation of quantum spin-dynamic effects with spin populations not restricted to the ground states. Our experiments show a temperature dependence that suggests the spin reversal undergoing a thermal instability that probably causes a magnetic deflagration. Molecular magnet Fe$_8$ has a biaxial anisotropy that brings interesting properties related to quantum tunneling of the magnetization. Thus observation of magnetic deflagration in the Fe$_8$ molecular magnet may open ahead new possibilities such is the observation of dipolar-field mediated deflagration \cite{Garanin2012}.

\section*{Acknowledgements}
F.M. acknowledges support from a Marie Curie IOF 253214 from EC and support from Catalan Government through COFUND-FP7. JT and FM also thank support from MAT2011-23698. The work at the KU Leuven is supported by the Methusalem Funding by the Flemish Government.

%\section*{Note}
%A recent experimental study seems to prove that magnetic deflagration occurs in Fe$_8$ \cite{keren_arxiv}. This manuscript also points towards how sensitive is the process to the sweep rate or to resonance conditions.

%\bibliography{biblio}

\end{document}